\documentstyle[12pt]{article}

\def\thefootnote{\fnsymbol{footnote}}
\renewcommand{\thefootnote}{\alph{footnote}}

\newcommand{\rref}[1]{(\ref{#1})}
\newcommand{\beqn}{\begin{equation}}
\newcommand{\eeqn}{\end{equation}}
\newcommand{\beqarr}{\begin{eqnarray}}
\newcommand{\eeqarr}{\end{eqnarray}}

\newcommand{\matc}{\begin{array}{c}}
\newcommand{\matcc}{\begin{array}{cc}}
\newcommand{\matccc}{\begin{array}{ccc}}
\newcommand{\matcccc}{\begin{array}{cccc}}
\newcommand{\emat}{\end{array}}

\newcommand{\Lag}{{\cal L}}

\begin{document}

\begin{titlepage}

July 1999         \hfill
\begin{center}
\hfill    UCB-PTH-99/26 \\
\hfill    LBNL-43404     \\

\vskip .15in
\renewcommand{\thefootnote}{\fnsymbol{footnote}}
{\large \bf Duality Invariant Born-Infeld Theory}
\vskip .30in

Daniel Brace\footnote{email address: brace@thsrv.lbl.gov}, 
Bogdan Morariu\footnote{email address: morariu@thsrv.lbl.gov} and 
Bruno Zumino\footnote{email address: zumino@thsrv.lbl.gov}
\vskip .30in

{\em    Department of Physics  \\
        University of California   \\
                                and     \\
        Theoretical Physics Group   \\
        Lawrence Berkeley National Laboratory  \\
        University of California   \\
        Berkeley, California 94720}
\end{center}
\vskip .40in

\begin{abstract}
We present an $Sp(2n,{\bf R})$
duality invariant  Born-Infeld $U(1)^{2n}$ 
gauge theory  with scalar fields.
To implement this duality we had to introduce complex gauge fields
and as a result the rank of the duality group is only half as large as 
that of the corresponding
Maxwell gauge theory with the same number of gauge fields. The latter
is self-dual under $Sp(4n,{\bf R})$, the largest allowed
duality group. A special case appears for $n=1$ when one can also write an
$SL(2,{\bf R})$ duality invariant Born-Infeld theory with a real gauge
field. We also describe the supersymmetric version of the
above construction. 
\end{abstract}

\end{titlepage}

\newpage
\renewcommand{\thepage}{\arabic{page}}

\setcounter{page}{1}
\setcounter{footnote}{0}


The general theory of duality invariance of abelian gauge theory 
developed in~\cite{GZ,BZ} was 
inspired by the
appearance of duality in extended supergravity theories~\cite{FSZ,CJ}.
For a theory with $M$ abelian gauge fields and appropriately chosen
scalars the maximal duality group is $Sp(2M,{\bf R})$.
The only known example with an $Sp(2M,{\bf R})$ duality symmetry,
where the Lagrangian is known in closed form is the 
Maxwell theory with $M$ gauge fields and an $M$-dimensional
symmetric matrix scalar field. 

Without scalar
fields the theory is self-dual only under the maximal compact subgroup 
of the duality group. Noncompact duality 
transformations relate theories at different values of the coupling constants.
The equations of motion derived from the
Born-Infeld Lagrangian with a  $U(1)$ gauge group 
are invariant under a $U(1)$ duality group just like pure 
electromagnetism~\cite{Sch}.
Introduction of scalar fields~\cite{GR1,GR2,GZ1,GZ2}, as described by 
the general theory developed in~\cite{GZ}, results in 
an $SL(2,{\bf R})$ duality 
invariant Born-Infeld theory. However, the proof of duality is 
complicated by the appearance of the square-root.

In this paper, inspired by the use of auxiliary fields in ~\cite{APT,RT},
we present an alternative form of the Born-Infeld action 
with scalars. The new form, without a square-root, 
simplifies the proof of duality invariance.
Furthermore, using this form we extend the Born-Infeld theory to
include  more than one 
abelian gauge field. However, these gauge fields must be complex. 
To obtain an $Sp(2n,{\bf R})$ duality group, the 
gauge group must be $U(1)^{2n}$.
The $n=1$ 
case is special in that both real and complex gauge fields are allowed.
For a single real gauge field, we  give both the 
formulation in terms of the auxiliary
fields and the square-root form obtained after eliminating the
auxiliary fields.

For the bosonic Born-Infeld with arbitrary $n$ we have calculated
the first few terms in the  square-root expansion and based on these
we conjecture the general form of the action without auxiliary
fields. It involves a symmetrized trace.
We also present an $N=1$ supersymmetrization of the
constructions described above.

We now briefly review the general theory of duality invariance
of an abelian gauge theory developed in~\cite{GZ}. 
However, we assume the gauge fields are complex, i.e. we start with an
even number of gauge fields and pair them into complex fields.
Consider an arbitrary Lagrangian 
\[
\Lag= \Lag(F^a,\bar{F}^a,\phi^i,\phi^i_{\mu}),
\]
where $\phi^i$ are some scalar fields, $F^{a}$ are $n$ complex field 
strengths, and $\bar{F}^a$ their complex conjugate . 
We define the dual field $G^a_{\mu\nu}$ or rather 
its Hodge dual $\tilde{G}^{a}_{\mu\nu}=\frac{1}{2}
\varepsilon_{\mu\nu\rho\sigma}G^{a\,\rho\sigma}$  as
\[
\tilde{G}^{a}_{\mu\nu}\equiv 2\frac{\partial \Lag}
{\partial  \bar{F}^{a\,\mu\nu}},~~
\tilde{\bar{G}}^{a}_{\mu\nu}\equiv 2\frac{\partial \Lag}
{\partial  F^{a\,\mu\nu}}.
\]

The main result of the paper~\cite{GZ}, and extended here to the case of
complex gauge fields, was to find the general 
conditions such that the equations of motion derived from the Lagrangian $\Lag$
are invariant under the infinitesimal transformations
\beqarr
\delta
\left(
\matc
G \\
F
\emat
\right)
&=&
\left(
\matcc
A & B \\
C & D
\emat
\right)
\left(
\matc
G \\
F
\emat
\right),
\label{FGtrans}  \\
\delta \phi^i &=& \xi^i(\phi).\nonumber
\eeqarr
In~\rref{FGtrans} we 
have combined the  field strengths $F$ and its dual $G$ 
into a $2n$-dimensional column vector, and $\xi^i$
are some unspecified transformations of the scalar fields.
The equations of motion are invariant if 
the matrices $A$, $B$, $C$, and $D$ are real and satisfy
\beqn
A^T= -D,~~B^T=B,~~C^T=C,
\label{ABrel}
\eeqn
and additionally the Lagrangian transforms as
\beqn
\delta \Lag= \frac{1}{2}(\bar{F} B\widetilde{F} +\bar{G}
C\widetilde{G})
,
\label{dLag}
\eeqn
where  all the space-time indices are contracted and a transposition 
with respect to the gauge index $a$ is used when necessary but is not
explicitly written.

The finite form of the transformation~\rref{FGtrans} is given by
\beqn
\left(
\matc
G' \\
F'
\emat
\right)=
\left(
\matcc
a & b \\
c & d
\emat
\right)
\left(
\matc
G \\
F
\emat
\right),
\label{Fintrans}
\eeqn
and must be an $Sp(2n,{\bf R})$ transformation.
This is the group of $2n$-dimensional matrices
where the $n$-dimensional blocks  $a$, $b$, $c$ and $d$  satisfy
\beqn
c^T a = a^T c,~~ b^T d = d^T b,~~ d^T a-b^T c =1.
\label{sp2n}
\eeqn
Using
\[
a\approx 1+A,~~b \approx B,~~c\approx C,~~d\approx 1+D,
\]
in~\rref{sp2n} and keeping only linear terms we obtain
the infinitesimal relations~\rref{ABrel}.  

One can check using~\rref{FGtrans} that the condition on the
variation of the Lagrangian~\rref{dLag}  is equivalent to the 
invariance of the following combination
\beqn
\Lag-\frac{1}{4}\bar{F} \widetilde{G}-\frac{1}{4} F
\widetilde{\bar{G}}.
\label{inv}
\eeqn
The linear combination~\rref{inv} must therefore also be invariant under finite
transformations. In the known examples for real gauge fields 
one can write the Lagrangian as
a sum of two pieces, the first invariant under $Sp(2n,{\bf R})$ and the 
second equal to  $\frac{1}{4}F\widetilde{G}$. Similarly, for complex
gauge fields there is an invariant piece and a piece equal to
$\frac{1}{4}\bar{F}\widetilde{G}+\frac{1}{4}F\widetilde{\bar{G}}$.

Now we are ready to describe the main result of this paper, a
Born-Infeld Lagrangian with a
$U(1)^{2n}$
gauge group, written in terms of the  auxiliary
fields, which is $Sp(2n,{\bf R})$ self-dual 
\beqn
\Lag={\rm Re}\left[{\rm Tr}\,(
i(\lambda-S)\chi -\frac{i}{2} \lambda \chi S_2 \chi^{\dagger}
+i\lambda(\alpha-i\beta) 
)\right].\label{BIL}
\eeqn
Here the auxiliary fields  $\lambda$ and $\chi$ are arbitrary complex 
$n$-dimensional matrices, $S$ is a complex symmetric matrix with a
positive definite imaginary part, and
$\alpha$ and $\beta$ are hermitian matrices defined in terms of
the complex gauge field strengths by
\[
\alpha^{ab} = \frac{1}{2} F^{a} \bar{F}^{b},
~~\beta^{ab} = 
\frac{1}{2}  \widetilde{F}^{a} \bar{F}^{b}.
\]
We can write the scalar fields as
\[  
S=S_1 + i S_2,~~\lambda=\lambda_1+i \lambda_2,~~\chi=\chi_1+i \chi_2,
\]
where $S_i$, $\lambda_i$ and $\chi_i$ are hermitian matrices. Note
that since $S$ is symmetric the $S_i$'s are real symmetric. The
Lagrangian~\rref{BIL} is invariant under a parity transformation, 
under which the fields
transform as
\beqarr
\alpha' &=& \bar{\alpha}, \nonumber  \\
\beta' &=& -\bar{\beta}, \nonumber   \\
S'        &=&   - \bar{S},  \label{parity}  \\
\chi'     &=&  \bar{\chi}, \nonumber \\
\lambda'  &=&  - \bar{\lambda}.  \nonumber 
\eeqarr 

The duality transformations of the scalar fields are given by
\beqarr
S'         &=&(aS+b)(cS+d)^{-1}, \nonumber \\
\lambda'   &=&(a\lambda+b)(c\lambda+d)^{-1},\label{duality}  \\
\chi'      &=&(c\lambda+d)\chi (cS+d)^{T}.\nonumber
\eeqarr
For convenience we also give the following transformation properties
derived from~\rref{duality}
\beqarr
S_2'         &=&(cS+d)^{-T} S_2(c\bar{S}+d)^{-1}, \nonumber \\
\lambda_2'   &=&(c\bar{\lambda}+d)^{-T}\lambda_2(c\lambda+d)^{-1},
\label{duality2}  \\
\chi'^{\dagger}      &=&(c\bar{S}+d)\chi^{\dagger} (c\bar{\lambda}+d)^{T}.
 \nonumber
\eeqarr
Explicit use of the symmetry of $S$ was used to obtain the first 
relation. 

Note that one can require a matrix transforming by
fractional transformation of the symplectic group to be symmetric. 
However the
transformation of $\chi$ is inconsistent with requiring
$\chi$ to be symmetric.  If we want
consistent equations of motions derived from the Lagrangian~\rref{BIL} 
we cannot require a symmetric $\lambda$ either. For $n\geq 2$ this in turn is
consistent with the transformations of the gauge field strengths and
their duals~\rref{Fintrans}  only if we take complex gauge fields.

The term  ${\rm Re}\left[{\rm Tr}\,(i\lambda(\alpha-i\beta) )\right]$ 
in the Born-Infeld Lagrangian~\rref{BIL} exactly cancels 
$\frac{1}{4} \bar{F}
\widetilde{G}+\frac{1}{4}F\widetilde{\bar{G}}$ 
in~\rref{inv}.
This is similar to the Maxwell theory where the noninvariant term 
can be written as $\frac{1}{4}F \widetilde{G}$.
The first two terms in the Born-Infeld Lagrangian~\rref{BIL} are
$Sp(2n,{\bf R})$ invariant.
To show the invariance it is convenient to use
\[
{\rm Re}\left[{\rm Tr}\,(-\frac{i}{2} \lambda \chi S_2
  \chi^{\dagger})\right]
={\rm Tr}\,(\frac{1}{2}\lambda_2 \chi S_2 \chi^{\dagger})
\]
to rewrite the second term, and then use the
transformations~\rref{duality} and~\rref{duality2}
to show the invariance 
of ${\rm Tr}\,(\frac{1}{2}\lambda_2 \chi S_2 \chi^{\dagger})$
and ${\rm Tr}\,(i(\lambda - S)\chi)$.

The equation of motion obtained by varying $\lambda$ is
\beqn
\chi-\frac{1}{2}\chi S_2 \chi^{\dagger} +\alpha-i\beta=0.
\label{EQ}
\eeqn
Using this in~\rref{BIL} the coefficient of $\lambda$ vanishes 
and the Lagrangian simplifies to
\beqn
\Lag={\rm Re}\left[{\rm Tr}\,(-i S\chi)\right]
={\rm Tr}\,(S_2\chi_1+S_1\chi_2).
\label{BI2}
\eeqn

For $n=1$ we can solve explicitly the equation~\rref{EQ}. It has two solutions
and we chose 
\[
\chi =\frac{1-\sqrt{1+2S_2\alpha - S_2^2 \beta^2}}{S_2}+i\beta,
\] 
such that the kinetic term in the action has the correct sign.
Using this in~\rref{BI2} we finally obtain
\beqn
\Lag=1-\sqrt{1+2 S_2\alpha - S_2^2 \beta^2} +S_1\beta.
\label{LBI}
\eeqn
As mentioned before, for $n=1$ we can also consider a single real gauge 
field. In this case we define $\alpha=\frac{1}{4} F^2$ 
and $\beta=\frac{1}{4}F\widetilde{F}$. 
If $\Lag_{BI}$ is the standard  Born-Infeld Lagrangian without scalar fields
\[
\Lag_{BI}(F)=1-\sqrt{\det(\eta+F)},
\]
we can also write~\rref{BIL} as
\beqn
\Lag=\Lag_{BI}(S_2^{1/2}F)+\frac{S_1}{4}
F\widetilde{F}.
\label{BI1}
\eeqn
As discussed in~\cite{GZ2}, this is the standard way of extending the
self-duality group from its compact form to the maximally split
noncompact duality group by introducing scalar fields.

For arbitrary $n$ the equation~\rref{EQ} implies
\[
\chi_2=\beta
\]
and using this in~\rref{EQ} we obtain an equation for $\chi_1$
\beqn
\chi_1=\frac{1}{2}
(\chi_1 S_2 \chi_1+ i\beta S_2\chi_1-i\chi_1 S_2 \beta +\beta S_2\beta)
-\alpha.
\label{chi1}
\eeqn
This equation simplifies using the following field redefinitions
\beqarr
\widehat{\chi_1}&=&S_2^{1/2}\chi_1 S_2^{1/2}, \nonumber \\
\widehat{\alpha}&=&S_2^{1/2}\alpha S_2^{1/2},  \\
\widehat{\beta}&=&S_2^{1/2}\beta S_2^{1/2}. \nonumber 
\eeqarr
The hatted variables satisfy equation~\rref{chi1} with $S_2=1$.
Then $X\equiv 1-\widehat{\chi_1}$ satisfies 
\beqn
X^2=1+2\widehat{\alpha}-\widehat{\beta}^2+i[\widehat{\beta},X],
\label{xEQ}
\eeqn
and the Lagrangian takes the form
\[
\Lag=n-{\rm Tr}\,(X)+{\rm Tr}\,(S_1\beta).
\]
We can solve the equation~\rref{xEQ} as an expansion in powers of $F^2$
\beqn
X=\sum_{m \geq 0} \frac{1}{m!} X^{m},
\label{Xexp}
\eeqn
where $m$ counts the number of times $F^2$ appears in $X^{m}$. One can
show using~\rref{xEQ} that $X^m$ satisfies the following recursion
relation
\[
2 X^m =-\sum_{j=1}^{m-1} \left(\matc m\\j\emat \right) X^j X^{m-j}+ A^m
+i m [\widehat{\beta},X^{m-1}],
\] 
where $X^0=1$, $A^1=2\widehat{\alpha}$, $A^2=-2\widehat{\beta}$ 
and $A^m=0$ for $m\geq 3$. 
We have calculated up to the ${\rm Tr} X^6$ term in the expansion for 
the trace of $X$ using~\rref{Xexp} and the result coincides 
with the symmetrized trace of the
square root of $1+2\widehat{\alpha}-\widehat{\beta}^2$
\beqn
{\rm Tr}\,(X)= {\rm Tr_{\cal S}}\,\sqrt{1+2\widehat{\alpha}-\widehat{\beta}^2}.
\label{noproof}
\eeqn
Here ${\rm Tr_{\cal S}}$ denotes the symmetrized trace where symmetrization
is with respect to $\widehat{\alpha}$ and $\widehat{\beta}$. One has
to first expand the
square root, then symmetrize each monomial in the expansion, and finally 
take the trace. We conjecture that the relation~\rref{noproof} is true 
to all orders. Then the Born-Infeld Lagrangian takes the form
\beqn
\Lag=n-{\rm Tr_{\cal S}}\,\sqrt{1+2\widehat{\alpha}-\widehat{\beta}^2}
+ {\rm Tr}\,{S_1\beta}.\label{xxx}
\eeqn
The appearance of only even powers of $\beta$ in  the second term
of~\rref{xxx}
is due to the
discrete symmetry~\rref{parity}.

In general one also adds to~\rref{BIL} a nonlinear
sigma model Lagrangian for the $S$ field. A metric invariant
under the transformations~\rref{duality} is given by
\beqn
{\rm Tr}\left[
(\bar{S}-S)^{-1}d \bar{S}(S-\bar{S})^{-1}d S
\right],
\label{metric}
\eeqn
which is a generalization of the metric on the Poincare upper half
plane. 

Finally we briefly discuss the supersymmetric Born-Infeld action.
Using the superfields $V^a=\frac{1}{\sqrt{2}}( V^a_1+iV^a_2)$ 
and  $\check{V}^a=\frac{1}{\sqrt{2}}(V^a_1-iV^a_2)$ 
where $V^a_1$ and $V^a_2$ are vector superfields,
we define
\[
W^a_{\alpha}=-\frac{1}{4}\bar{D}^2 D_{\alpha} V^a,~~
 \check{W}^a_{\alpha}=-\frac{1}{4}\bar{D}^2 D_{\alpha} \check{V}^a.
\]
Note that both $W^a$ and $\check{W}^a$ are chiral superfields. Let us
also define
\[
{\cal M}^{ab}= W^a \check{W}^b.
\]
We can construct the supersymmetric version of the Lagrangian~\rref{BIL}
\beqn
\Lag={\rm Re}
\int \, d^2 \theta
\left[{\rm Tr}\,(
i(\lambda-S)\chi -\frac{i}{2} \lambda \bar{D}^2( \chi S_2 \chi^{\dagger})
- i  \lambda{\cal M}
)\right].\nonumber
\eeqn
Here $S$, $\lambda$ and $\chi$ are chiral superfields with the same
symmetry properties as the corresponding bosonic fields.
The bosonic fields $S$ and $\lambda$ appearing in~\rref{BIL} are 
the lowest component of the superfields denoted by the same letter. 
The field $\chi$  in the action~\rref{BIL} is the
highest component of the superfield $\chi$.

Just as in the bosonic case for $n=1$ we can also consider a
Lagrangian with a single real superfield. In this case one can
integrate out the auxiliary superfields and obtain a supersymmetric
version of~\rref{BI1}  
\beqn
\Lag=\int \, d^4 \theta\frac{S_2^2 W^2\bar{W}^2}{1-A+\sqrt{1-2A+B^2}}
+
{\rm Re}\left[
\int \, d^2 \theta (-\frac{i}{2} S W^2)\right],
\label{sbi}
\eeqn
where
\[
A=\frac{1}{4}(D^2(S_2W^2)+\bar{D}^2(S_2\bar{W}^2)),~~
B=\frac{1}{4}(D^2(S_2W^2)-\bar{D}^2(S_2\bar{W}^2)).
\]
For $S=i$ this reduces to the supersymmetric Born-Infeld action
described in~\cite{DP,CF,BG}. In the case of weak fields the
first term of~\rref{sbi} can be neglected and the Lagrangian is
quadratic in the field strengths. Under these conditions 
the combined requirements of
supersymmetry and self-duality have been used in~\cite{BG2} to constrain
the form of the weak coupling limit of effective supergravity 
Lagrangians describing the low energy limit of
string theory.

\section*{Acknowledgments}
We would like to dedicate this paper to the memory of Yuri Golfand,
one of the pioneers of supersymmetry.

We are indebted to P. Aschieri and especially to M. K. Gaillard 
for many illuminating discussions.
This work was supported in part by the Director, Office of Science,
Office of High Energy and Nuclear Physics, Division of High Energy
Physics of the U.S. Department of Energy under Contract
DE-AC03-76SF00098 and in part by the National Science Foundation 
under grant PHY-95-14797.

\end{document}